\documentstyle{l-aa}
\input psfig.sty

\begin{document}

\thesaurus{10(03.13.6; 04.19.1; 10.08.1; 10.11.1; 10.19.2; 10.19.3)}

\title{New surveys of UBV photometry and absolute proper 
motions at intermediate latitude\thanks{Based on observations made on photographic plates 
obtained with the Tautenburg, ESO and Palomar Schmidt telescopes and on
CCD observations made at UPSO, Nainital (India) and OHP, France. 
Digitizations made with the MAMA measuring machine. MAMA is developed 
and operated by INSU (Institut National des Sciences de l'Univers, Paris).
}  
}

\author{D.K. Ojha \inst{1}
\and O. Bienaym\'e \inst{2}
\and V. Mohan \inst{3}
\and A.C. Robin \inst{2,4}
}
\offprints{D. K. Ojha (ojha@tifr.res.in)}

\institute{Tata Institute of Fundamental Research, Homi Bhabha Road, Colaba, 
Mumbai - 400 005, India
\and
Observatoire de Strasbourg, CNRS URA 1280, 11 rue de l'Universit\'e, 
F-67000, Strasbourg, France
\and
Uttar Pradesh State Observatory, Manora Peak, Nainital, 263 129, India
\and
Observatoire de Besan\c con, 41 bis, Av. de l'Observatoire, BP 1615,
F-25010 Besan\c con Cedex, France
}

\date{Received xxxx, 1999, accepted xxxx, 1999}

\maketitle

\markboth {D.K.Ojha et al. : New surveys of UBV photometry and absolute 
proper motions at intermediate latitude}
{D.K.Ojha et al. : New surveys of UBV photometry and absolute proper 
motions at intermediate latitude} 

\begin{abstract}

A photometric and proper motion survey has been obtained in 2 directions
at intermediate latitude: ($l=167.5^\circ$, $b=47.4^\circ$; 
$\alpha_{2000}=9^h41^m26^s$,$\delta_{2000}=+49^\circ53'27''$) and
($l=278^\circ$, $b=47^\circ$; $\alpha_{2000}=11^h42^m56^s$,
$\delta_{2000}=-12^\circ31'42''$). The survey covers 7.13 and 20.84
square degrees, respectively. The limiting magnitude is about 18.5 in V
for both directions. 

We have derived the density laws for stars (M$_{V}$ $\ge$ 3.5) as a 
function of distance from the galactic plane. The density laws for 
stars follow a sum of two exponentials with scale heights of 240 pc
(thin disk) and 790 pc (thick disk), respectively. The local density 
of thick disk is found to be 6.1$\pm$3 \% relative to the thin disk. 
The kinematical 
distribution of stars has been probed to distances up to 3.5 kpc above
the galactic plane. New estimates of the parameters of velocity
ellipsoid have been derived for the thick disk of the Galaxy.
A comparison of our data sets with the Besan\c con model star count 
predictions has been performed, giving a good agreement in the magnitude 
range V = 13 to 18. 


\end{abstract}

\section {Introduction}

We have developed a programme of star counts on Schmidt plates in a few
directions of the Galaxy. Objectives of this programme are to trace the fine 
structure of our Galaxy through the statistical study of the stellar 
distributions according to their luminosity, colors and proper motions. This
involves two steps : first, acquiring a new photometric and 
astrometric sample survey in various galactic directions; and 
second analysing the data using a model of population synthesis
and determining the properties of populations in the Galaxy and 
constraints on the scenario of formation and evolution. In the framework
of this programme, new observations were carried out in selected areas
situated in the main meridional section and towards galactic antirotation
direction in the Galaxy. The combination of OCA, ESO, Tautenburg and
Palomar Schmidt plates is used to derive the photometry and the absolute
proper motions of the stars on a time baseline of about 35 years. 

In our earlier work (Ojha et al. 1994ab; Robin et al. 1996), using the data 
from galactic anticentre and centre fields with wide-area surveys in other 
fields available to date, we discussed the radial and vertical structure of
the Galaxy. By comparing the star count ratio between the two data sets
(galactic anticentre and centre), we obtained the scale length of thin disk
(h$_{R}$$\sim$2.5 kpc) and thick disk populations (h$_{R}$$\sim$3 kpc). 
The thick disk was found to have a scale height
of 760$\pm$50 pc, with a local density of 5.6$\pm$1 \% of the thin disk.
Our results confirmed no radial and vertical gradients in thick disk 
kinematics. The thick disk formation scenario was discussed based on the
new photometric and kinematic data.

The present paper extends our previous work of a sample survey plan to
produce probes of stellar populations in the Milky Way. The thick disk
population is revisited under the light of new data. We have chosen a
direction towards galactic anticentre at medium latitude
($l=167.5^\circ$, $b=47.4^\circ$;  
$\alpha_{2000}=9^h41^m26^s$,$\delta_{2000}=+49^\circ53'27''$; hereafter GAC2)
and obtained BV
star counts and absolute proper motion data in 7.13 square degrees for
a complete sample of 6041 stars brighter than B = 20 mag. Schmidt plates from
Tautenburg telescope were used to obtain magnitudes and color.
Deep CCD sequences from 104-cm telescope of
Uttar Pradesh State Observatory, Nainital (India) and 120-cm
telescope from Observatoire de Haute-Provence (France) have been
used to increase the magnitude limits up to V$\sim$18.5 and
B$\sim$20. The choice of this particular area was made partly
because this field gives the counterpart of the galactic centre field
(Ojha et al. 1994b, hereafter GC, Paper II) and serves as a complement 
to the field toward the outer part of the Galaxy. The absolute proper motions 
of the stars in this field have been published elsewhere 
(Ojha et al. 1994a, hereafter GAC1, Paper I).
Using the two fields (GAC1,2 \& GC) together is a good
choice for sighting the intermediate population as shown in our
preliminary studies (Paper I \& II; Ojha et al. 1996 (hereafter Paper III); 
Robin et al. 1996).

The other survey is based on the measurement of 14 Palomar and ESO Schmidt 
plates covering an area of 20.84 square degrees centered on 
$\alpha_{2000}=11^h42^m56^s$,$\delta_{2000}=-12^\circ31'42''$ 
corresponding to $l=278^\circ$, $b=47^\circ$. This field is close to
the galactic antirotation direction (hereafter GAR) at intermediate 
latitude. The catalogue contains about 22000 stars with 
completeness limits of V=18.5, B=20 and U=18.5. This field is a test 
direction to find evidence 
of the relative motions of the halo and intermediate populations. 
It also gives informations on the inclination of the velocity ellipsoids of 
these populations. These results will be published in future papers.
A preliminary analysis of this catalogue together
with a comparison with existing catalogues towards the same latitude 
($\mid$b$\mid$) and with Besan\c con model of population synthesis 
is given elsewhere (Paper III \& Robin et al. 1996). In the present paper,
we discuss a new complete survey of absolute stellar proper motions and
multicolor photometery. The more detailed analysis is also presented in the 
paper.

The outline of the paper is as follows: In \S 2, we discuss briefly the
photographic plates used in the present surveys. The photographic
calibration is explained in \S 3. In this section, we give the new
color equations for the ESO and Tautenburg Schmidt plates. 
\S 4 describes the derivation of absolute proper motions and the 
overall accuracy of the proper motion surveys. In \S 5, we present the
variation of the stellar number density as a function of distance from
the galactic plane. From this, we deduce the structural parameters of
the thin and thick disk population. In \S 6, we derive the kinematics
of the thick disk component. \S 7 shows the comparison of our data sets with
the Besan\c con model of population synthesis.
    
\section {Photographic material}

In the GAC2 field, among the 14 Schmidt plates used for the astrometric
reduction described in Paper I (Table 1), 
we have chosen 2 Tautenburg Schmidt plates (2420 \& 2430) in
the V band (Kodak 103a-G+GG11) and 2 plates (6568 \& 6569)
in the B band (Astro-Spezial+GG13) for photometric purposes. 
Each plate covers an area of 3.$^\circ$3$\times$3.$^\circ$3 with a scale
of 51.4 arcsec mm$^{-1}$. 
All the plates were measured on the MAMA
microdensitometer in Paris with a pixel size of 10 $\mu m$.
The treatment of the plate with MAMA has been extensively described
by Berger et al. (1991).

In the GAR field, 14 Schmidt plates from Palomar and ESO  
were used for the photometric and astrometric measurements. Plate
descriptions are given in table 1. 

\begin{table*}
\caption[]{Plate material used for GAR field}
\begin{flushleft}
\normalsize
\begin{tabular}{c c c c c}\\
\hline
Plate   &  Emulsion +  &  Color  & Epoch & Scale (''/mm)\\
number  &  filter \\
\hline
\it Palomar Schmidt plates\\
POSS 1562 & 103a-E+red plexiglass & red & 10/03/1956 & 67.14\\
POSS 1562 & 103a-O+blue plexiglass & blue & 10/03/1956 & 67.14\\
\it ESO Schmidt plates\\
8520 & IIaO+GG385  & B & 13/04/1990 & 67.13\\
8522 & IIaO+GG385  & B & 15/04/1990 & 67.13\\
8551 & 103aD+GG495 & V & 18/05/1990 & 67.13\\
8557 & 103aD+GG495 & V & 19/05/1990 & 67.13\\
8556 & IIaO+UG1    & U & 19/05/1990 & 67.13\\
8561 & IIaO+UG1    & U & 20/05/1990 & 67.13\\
9172 & IIaO+GG385  & B & 05/04/1991 & 67.13\\
9174 & IIaO+GG385  & B & 06/04/1991 & 67.13\\
9193 & IIIaF+RG630 & R & 12/04/1991 & 67.13\\
9194 & IIIaF+RG630 & R & 12/04/1991 & 67.13\\
9246 & 103aD+GG495 & V & 05/05/1991 & 67.13\\
9254 & IIIaF+RG630 & R & 05/05/1991 & 67.13\\
\hline
\end{tabular}
\end{flushleft}
\end{table*}

\section {Photometric reduction}

\subsection {Photometric sequences}

To photometrically calibrate the plates, a number of photometric 
standards are required. The standards should cover the entire 
range of magnitudes to be studied and should also be spread over 
the surface of the plate so as to minimise the geometrical effects 
present on the plate. To obtain the photometric standards, we have observed 
a number of subfields of each plate using the CCD system attached to 
the 1-metre telescope of the U. P. State Observatory at Nainital, 
India. A few standards have also been observed using the 1.2-metre 
telescope of Observatoire de Haute-Provence in France. The CCD images 
have been obtained in UBV filters. The image processing of the 
CCD images has been done using the ESO MIDAS and DAOPHOT softwares.

\subsection {GAC2}

Photographic BV magnitudes were derived from the total intensity using
the BV photometric sequences for the present survey.
The CCD magnitude sequences for this field were observed by
Mohan et al. (1994) in the U, B, and V passbands. The BV photometric
sequences include 110 stars for 11$<$V$<$21.
A fourth-order polynomial function is fitted to these data to give
the calibration curve for each plate.
The rms scatters of the fitting are in a range of 0.02 to 0.06
magnitude in the V and B bands. The color equations to convert
the instrumental system to the Johnson one have been derived
by the procedures described by Mohan \& Cr\'ez\'e (1987) and 
Ojha (1994). The
following new color equations are derived for the Tautenburg
Schmidt plates:

$$ v_{inst} = V_{John}+0.117(B-V) \eqno(1)$$
$$ b_{inst} = B_{John}+0.113(B-V) \eqno(2)$$

The plate to plate dispersion in all magnitude ranges is 
shown in table 2.

\begin{table}
\caption[]{Dispersion of magnitudes from plate to plate comparison 
in case of GAC2 field}
\begin{flushleft}
\normalsize
\begin{tabular}{c c c c}\\
\hline
V & $\sigma_{V}$ & B & $\sigma_{B}$\\
\hline
11.75 & 0.02 & 11.25 & 0.02 \\
12.75 & 0.03 & 12.25 & 0.02 \\
13.75 & 0.03 & 12.75 & 0.03 \\
14.25 & 0.04 & 13.25 & 0.03 \\
14.75 & 0.04 & 13.75 & 0.02 \\
15.25 & 0.04 & 14.25 & 0.03 \\
15.75 & 0.04 & 14.75 & 0.02 \\ 
16.25 & 0.05 & 15.25 & 0.03 \\
16.75 & 0.05 & 15.75 & 0.02 \\
17.25 & 0.05 & 16.25 & 0.03 \\
17.75 & 0.05 & 16.75 & 0.03 \\
18.25 & 0.06 & 17.25 & 0.03 \\
      &      & 17.75 & 0.03 \\
      &      & 18.25 & 0.04 \\
      &      & 18.75 & 0.04 \\
      &      & 19.25 & 0.05 \\
      &      & 19.75 & 0.06 \\
\hline
\end{tabular}
\end{flushleft}
\end{table}

\subsection{GAR}

Photometry is made using at least two plates per color in the
three photometric bands UBV. Photometric sequences used are
CCD photometric ones by Mohan et al. (1996).
About 135 stars are typically used to determine the calibration
curve in each passband. New color transformations between the
Johnson system and the instrumental system of ESO Schmidt plates 
have been determined. The color equations are given as follows :

$$ v_{inst} = V_{Johnson} - 0.130(B-V)\eqno(3) $$
$$ b_{inst} = B_{Johnson} - 0.143(B-V) \eqno(4)$$
$$ u_{inst} = U_{Johnson} + 0.245(B-V) - 0.056(U-B) \eqno(5)$$

The typical rms magnitude scatter ranges from 0.03 to 0.12 in the
magnitude range 11 to 19. 
The plate to plate dispersion in all magnitude ranges is shown in 
table 3.

\begin{table}
\caption[]{Dispersion of magnitudes from plate to plate comparison 
in case of GAR field}
\begin{flushleft}
\normalsize
\begin{tabular}{c c c c c c c c}\\
\hline
V & $\sigma_{V}$ & B & $\sigma_{B}$ & U & $\sigma_{U}$ \\
\hline
11.25 & 0.06 & 11.25 & 0.06 & 11.25 & 0.03 \\
11.75 & 0.06 & 11.75 & 0.06 & 11.75 & 0.05 \\
12.25 & 0.06 & 12.25 & 0.06 & 12.25 & 0.04 \\
12.75 & 0.06 & 12.75 & 0.06 & 12.75 & 0.04 \\
13.25 & 0.07 & 13.25 & 0.07 & 13.25 & 0.04 \\
13.75 & 0.06 & 13.75 & 0.06 & 13.75 & 0.04 \\
14.25 & 0.07 & 14.25 & 0.06 & 14.25 & 0.05 \\ 
14.75 & 0.07 & 14.75 & 0.06 & 14.75 & 0.05 \\
15.25 & 0.07 & 15.25 & 0.06 & 15.25 & 0.05 \\
15.75 & 0.08 & 15.75 & 0.06 & 15.75 & 0.05 \\
16.25 & 0.08 & 16.25 & 0.07 & 16.25 & 0.06 \\
16.75 & 0.09 & 16.75 & 0.07 & 16.75 & 0.06 \\
17.25 & 0.09 & 17.25 & 0.07 & 17.25 & 0.07 \\
17.75 & 0.10 & 17.75 & 0.08 & 17.75 & 0.08 \\
18.25 & 0.12 & 18.25 & 0.09 & 18.25 & 0.09 \\
      &      & 18.75 & 0.09 & 18.75 & 0.10\\
      &      & 19.25 & 0.10 \\
\hline
\end{tabular}
\end{flushleft}
\end{table}

\subsection{Star-galaxy separation}

It is important to separate the galaxies from stars in the 
catalogue.
The method of star/galaxy separation
is described in our previous paper (Paper I). The classification
of an image as a star or a galaxy was determined from the deepest and 
reference plate, V8557 in case of GAR field. 
Because the galaxy sequences merge with the
stellar sequences at the faint end, this has the consequences that 
for V$>$18.5 the inclusion of galaxies in star bin is very high. For
V$\le$ 18, the discrimination between stars and galaxies is very
good and there is essentially no contamination of the star counts 
by galaxies (see also \S 7). Similarly,  
Tautenburg plate 6568 was used to separate stellar and
non-stellar objects in case of GAC2 field. The star-galaxy discrimination is
reliable to V $\sim$ 18 mag.

\subsection{Completeness}

The instrumental photographic magnitudes are converted to
the Johnson system using Equs. (1) to (5).
Histogram of standard V magnitude counts from GAC2 and GAR field are shown
in Fig. 1. The completeness limits of the U, B and V counts are
found to be 18.5, 20 and 18.5 in two fields.

\begin{figure*}
\psfig{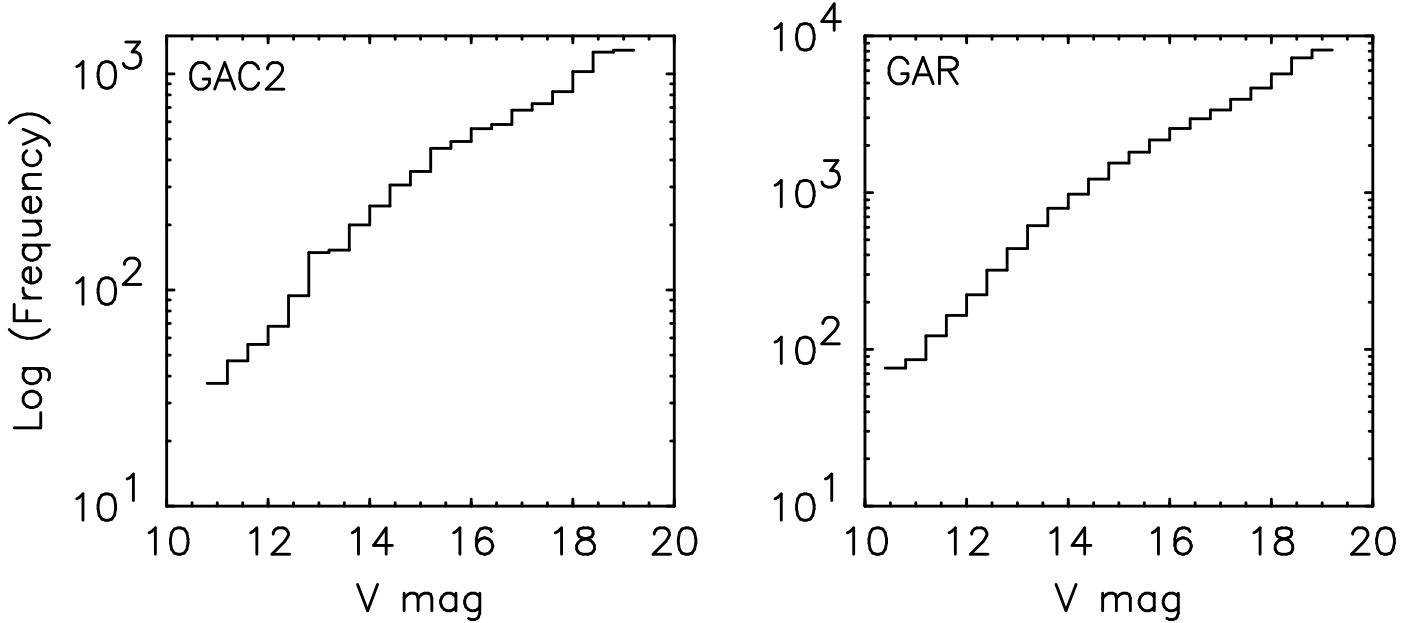}
\caption[]{Histograms of V star counts in GAC2 and GAR field. The completeness 
limit is V = 18.5}
\end{figure*}

\subsection{Photometric catalogues}

Our final photometric catalogues include 6041
stars over 7.13 square degrees with BV magnitude and proper motions in GAC2
field and 22000 stars over 20.84 square degrees with UBV magnitude and proper
motions in GAR field, respectively. Since both fields are at intermediate
latitudes ($\mid$b=47$^\circ$$\mid$), therefore the interstellar reddening is 
negligible, hence no correction has been applied to unreddened the main
sequence.
The observed starcount data are also presented in a tabular form in
Tables 4, 5 and 6 for N(V,B-V) and N(V,U-B). The color-color diagrams of the
sources from two fields are shown in Fig. 2. The full line shown in 
the figures locates the main sequence.

\begin{table*}
\caption[]{Starcounts over 7.13 square degrees as a function of V and B-V 
(GAC2 field)}
\begin{flushleft}
\normalsize
\begin{tabular}{c c c c c c c c c c c c c c c c}\\
\hline
B-V&-0.4&-0.2&0.0&0.2&0.4&0.6&0.8&1.0&1.2&1.4&1.6&1.8&2.0&2.2&Total\\
V\\
\hline
11.0-11.5 & 0 & 0& 1 &8 &10 &12 &9 &9 &0 &0 & 0&0 &0 &0 &49\\
11.5-12.0 & 0 & 0& 0 &5 &18 &12 &10 &9 &4 &0 & 0&0 &0 &0 &58\\
12.0-12.5 & 0 & 0& 0 &1 &26 &23 &12 &8 &0 &0 & 0&0 &0 &0 &70\\
12.5-13.0 & 0 & 0& 0 &2 &23 &63 &27 &14 &0 &1 & 0&0 &0 &0 &130\\
13.0-13.5 & 0 & 0& 0 &3 &29 &72 &42 &11 &4 &3 & 0&0 &0 &0 &164\\
13.5-14.0 & 0 & 0& 0 &3 &27 &93 &47 &15 &3 &3 & 0&0 &0 &1 &191\\
14.0-14.5 & 0 & 0& 0 &1 &36 &130 &56 &12 &8 &8 & 0&0 &0 &0 &251\\
14.5-15.0 & 0 & 0& 0 &1 &20 &153 &98 &40 &17 &4 & 2&0 &0 &0 &335\\
15.0-15.5 & 0 & 0& 0 &1 &34 &162 &123 &55 &24 &17 &3&2 &0 &0 &421\\
15.5-16.0 & 0 & 0& 0 &2 &26 &174 &151 &82 &55 &30 &12&0 &2 &0 &534\\
16.0-16.5 & 0 & 1& 1 &1 &27 &169 &161 &97 &59 &41 &23&4 &0 &0 &584\\
16.5-17.0 & 0 & 0& 2 &5 &37 &139 &135 &123 &87 &69 &39&6 &1 &0 &643\\
17.0-17.5 & 0 & 0& 0 &7 &25 &134 &142 &136 &109 &100 &72&7 &4 &0 &736\\
17.5-18.0 & 0 & 1& 0 &4 &55 &132 &139 &125 &126 &153 &113&13 &4 &2 &867\\
18.0-18.5 & 1 & 2& 5 &11&66 &112 &131 &143 &164 &211 &147&11 &0 &0 &1004\\
\hline
\end{tabular}
\end{flushleft}
\end{table*}

\begin{table*}
\caption[]{Starcounts over 20.84 square degrees as a function of V and B-V
(GAR field)}
\begin{flushleft}
\normalsize
\begin{tabular}{c c c c c c c c c c c c c c c c}\\
\hline
B-V&-0.4&-0.2&0.0&0.2&0.4&0.6&0.8&1.0&1.2&1.4&1.6&1.8&2.0&2.2&Total\\
V\\
\hline
11.0-11.5& 0 &  0 &  6 & 30 & 21 &  9 &19  &8  &8  &0  &0  &0 &0&0&101\\
11.5-12.0& 0 &  0 &  0 & 13 & 54 & 33 &23  &13 &1  &1  &2  &0 &0&0&140\\
12.0-12.5& 0 &  0 &  1 & 12 & 69 & 57 &42  &24 &4  &2  &0  &0 &0&0&212\\
12.5-13.0& 0 &  0 &  1 &  5 & 93 &136 &68  &22 &11 &2  &1  &0 &0&0&338\\
13.0-13.5& 0 &  0 &  1 &  5 &110 &245 &115 &46 &9  &8  &0  &0 &0&0&539\\
13.5-14.0& 0 &  1 &  3 &  4 & 90 &357 &186 &89 &22 &6  &0  &0 &0&0&760\\
14.0-14.5& 0 &  1 &  4 &  3 & 56 &427 &299 &108&37 &15 &2  &2 &0&0&954\\
14.5-15.0& 0 &  0 &  3 &  6 & 66 &511 &422 &166&69 &34 &10 &1 &1&0&1289\\
15.0-15.5& 0 &  1 &  3 &  5 & 83 &580 &547 &221&83 &59 &25 &1 &3&0&1611\\
15.5-16.0& 0 &  1 &  2 & 13 &103 &599 &678 &327&147&115&63 &6 &0&1&2056\\
16.0-16.5& 0 &  2 & 12 & 17 &143 &690 &686 &374&182&139&107&21&3&0&2379\\
16.5-17.0& 0 &  1 &  5 & 18 &182 &788 &690 &366&282&202&204&40&4&3&2786\\
17.0-17.5& 1 &  2 & 12 & 34 &303 &843 &708 &427&277&298&296&68&7&0&3276\\
17.5-18.0& 0 &  2 & 15 & 75 &455 &853 &677 &443&373&405&341&39&1&0&3677\\
18.0-18.5& 2 &  7 & 32 & 128&616 &964 &714 &494&273&105&0  &0 &0&0&3337\\
\hline
\end{tabular}
\end{flushleft}
\end{table*}

\begin{table*}
\caption[]{Starcounts over 20.84 square degrees as a function of V and U-B
(GAR field)}
\begin{flushleft}
\normalsize
\begin{tabular}{c c c c c c c c c c c c c c c c c c c}\\
\hline
U-B&-1.0&-0.8&-0.6&-0.4&-0.2&0.0&0.2&0.4&0.6&0.8&1.0&1.2&1.4&1.6&
1.8&2.0&2.2&Total\\
V\\
\hline
11.0-11.5&0&0&5&10&17&8&6&7&8&5&1&2&3&2&0&0&0&74\\
11.5-12.0&0&2&13& 30& 28& 10& 15&  5&  9&  7& 3& 1& 0&1&1&0&0&125\\
12.0-12.5&0&4&18& 37& 45& 20& 16& 15& 13& 16& 4& 5& 1&0&1&0&0&195\\
12.5-13.0&0&0&27& 58& 77& 46& 27& 27& 16&  8&11& 4& 1&4&1&0&0&307\\
13.0-13.5&0&0&23&105&116& 96& 53& 39& 29& 20& 9& 4& 1&1&0&0&0&496\\
13.5-14.0&0&0&13& 99&160&150& 93& 75& 46& 38&20& 7& 3&0&0&0&0&704\\
14.0-14.5&0&0& 4& 68&167&216&157&114& 70& 38&34&21& 4&3&0&0&0&896\\
14.5-15.0&0&1& 1& 42&220&317&210&160&101& 47&50&51&12&0&0&1&0&1213\\
15.0-15.5&0&0& 1& 33&207&376&296&209&137&101&63&58& 7&0&1&0&0&1489\\
15.5-16.0&1&0& 0& 18&181&445&366&291&191&121&77& 9& 0&0&0&0&0&1700\\
16.0-16.5&0&2& 1& 18&240&522&404&277&187& 62& 5& 0& 0&0&0&0&0&1718\\
16.5-17.0&1&2& 2& 15&246&649&437&155& 20&  0& 0& 0& 0&0&0&0&0&1527\\
17.0-17.5&0&1& 3& 18&335&533&135&  1&  0&  0& 0& 0& 0&0&0&0&0&1026\\
17.5-18.0&1&6& 6& 36&174& 48&  3&  0&  0&  0& 0& 0& 0&0&0&0&0&274\\
18.0-18.5&7&5&15&  4&  1&  0&  0&  0&  0&  0& 0& 0& 0&0&0&0&0&32\\
\hline
\end{tabular}
\end{flushleft}
\end{table*}

\begin{figure*}
\psfig{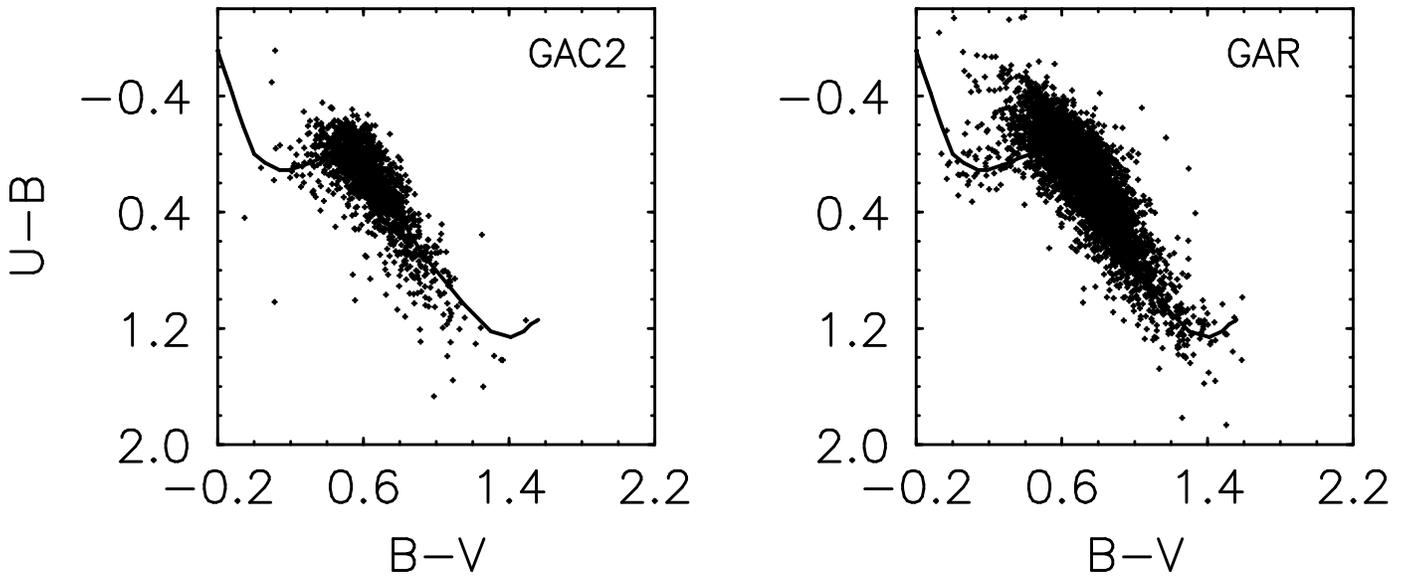}
\caption[]{U-B {\it versus} B-V diagrams of stars in GAC2 and GAR field. 
The full line locates the main sequence}
\end{figure*}

\section{Astrometry}

For the astrometric reduction,
the displacement of each stellar image is measured relative to the 
framework defined by all stars in the field. Then the relative 
proper motions have been shifted to absolute proper motions using 
the extragalactic objects in the field. Orthogonal functions were 
used to model transform between the plate coordinates. The distribution 
of reference stars is used to define an orthogonal system. This method 
is described in Bienaym\'e (1993) and in Paper I \& II.
 
\subsection{Absolute proper motions}

For GAR field, 
the mean proper motions of a sample of $\sim$ 3842 galaxies 
uniformly distributed in color (B-V$<$2.5) and magnitude ranges 
(12.5$<$V$<$18.5) has
been used to calculate the zero point of the proper motions.
The conversion equations obtained are as follows:
\vskip 0.4cm
$\mu_{\alpha}(abs) = \mu_{\alpha} - 0.65\pm0.02$~~(''/cen)
\vskip 0.4cm
$\mu_{\delta}(abs) = \mu_{\delta} - 0.27\pm0.01$~~(''/cen)

The mean error of the differential proper motions in 
arcsec per century as a function of V magnitude is given 
in table 7.

\begin{table}
\caption[]{The mean error (arcsec per century) in proper
motion $<$ $\sigma_{\mu}$ $>$ = $\sqrt{\sigma_{\mu_{x}}^{2}+
\sigma_{\mu_{y}}^{2}}$ as a function of V magnitude in case of 
GAR field}
\begin{flushleft}
\normalsize
\begin{tabular}{c c}\\
\hline
V mag & $<$$\sigma_{\mu}$$>$\\
interval& (''/cen)\\
\hline
11.25 & 0.22\\
11.75 & 0.22\\
12.25 & 0.23\\
12.75 & 0.22\\
13.25 & 0.21\\
13.75 & 0.19\\
14.25 & 0.20\\
14.75 & 0.18\\
15.25 & 0.14\\
15.75 & 0.15\\
16.25 & 0.12\\
16.75 & 0.12\\
17.25 & 0.12\\
17.75 & 0.13\\
18.25 & 0.17\\
\hline
\end{tabular}
\end{flushleft}
\end{table}

The zero points of the proper motions for GAC2 field are given elsewhere 
(Paper I). In Fig. 3, the Vector Point Diagrams of the
sources in two fields are presented. The asymmetric distribution in 
figures is consequence of the stellar asymmetric drift because of the
growing contribution from higher velocity stars.

\begin{figure*}
\psfig{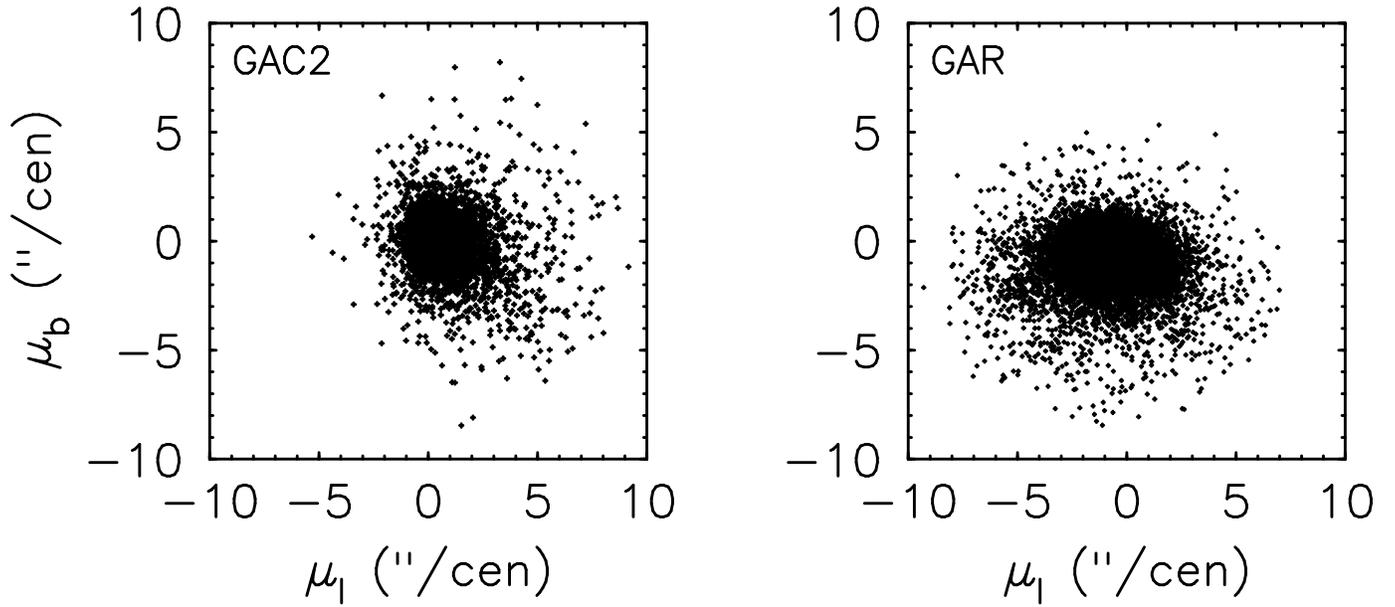}
\caption[]{Vector point diagram in GAC2 and GAR field}
\end{figure*}

\section{Space density of stars in GAR direction}

\subsection{Stellar distance}

We have used BV photographic photometry to derive the distances of
the sub-sample of stars. The distances were determined by estimating 
absolute stellar magnitudes, which were obtained from a M$_{V}$ $versus$
B-V relation, taking into account the metallicity change as a function
of the distance from the galactic plane.
The derivation of the stellar distance estimates
is explained in Paper III. We assume a vertical gradient of
metallicity ($\partial$ [Fe/H]/$\partial$z = -0.3 kpc$^{-1}$) to correct the 
absolute magnitude (Kuijken \& Gilmore 1989).

\subsection{Density laws}

We have derived the logarithmic space density functions of the
thin and thick disk stars in 3.5$\le$M$_{V}$$\le$4.5 and 
4.5$\le$M$_{V}$$\le$6.0 absolute magnitude intervals. The 
derivation for the space density of stars is explained in detail
in Paper III. The observed density is fitted by the sum
of two exponentials of the form shown in Eq. (1) in Paper III.
The values of the scale lengths of thin and thick disk populations
determined from the star count ratio (Paper III) were fixed
in the calaculation.

A least-squares method is adopted to derive
the best model in which the exponential distributions of the
number density reproduce the observed starcounts. In Fig. 4,
we present the density distribution of stars for two different
absolute magnitude ranges, as a function of distance above the galactic
plane. Table 8 presents the best fitted structural parameters
for stars with M$_{V}$ $\ge$ 3.5 in GAR field. In the same table, we
have also presented the mean structural parameters from the other two  
intermediate latitude fields (galactic anticentre \& centre),
which give an idea of the error on the determination.  
 
\begin{figure}
\psfig{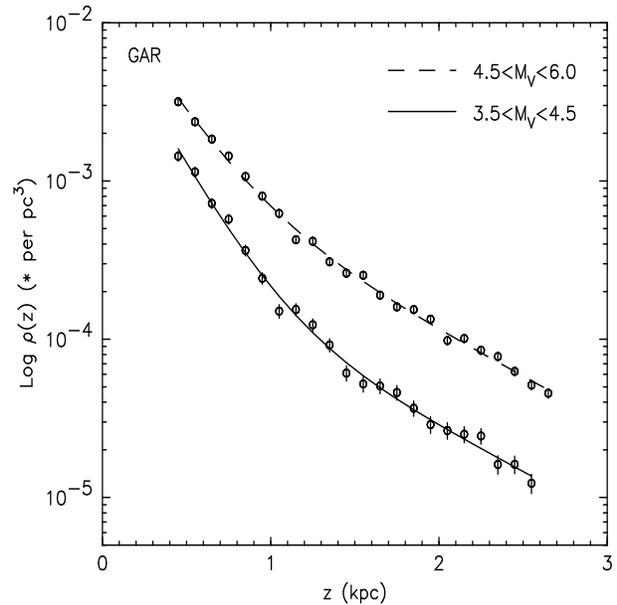}
\caption[]{The density distribution for stars with M$_{V}$ $\ge$ 3.5
as a function of distance above the galactic plane. The fitted lines
represent the sum of two exponentials with scale heights given in
table 8}
\end{figure}

The value of thin disk scale height (h$_{z}$ $\sim$ 240 pc) derived from
GAR field is in 
agreement with the values obtained by Kuijken \& Gilmore (1989) from a 
K dwarf photometric parallax study in the direction of the South Galactic
Pole. Ng et al. (1995) determination gives h$_{z}$ = 250 pc based on a sample
of stars towards the galactic centre. Haywood (1994, 1997) showed that the
overall vertical density profile of the galactic disk is closed to an
exponential with scale height h$_{z}$$\simeq$250 pc. 

Our value of thick disk scale height derived from GAR field is in good 
agreement with the one derived from two intermediate latitude fields at 
galactic centre and anticentre (Paper III \& table 8). Robin et al. (1996) 
determination 
gives h$_{z}$ = 760$\pm$50 pc, with a local density of 5.6$\pm$1 \% relative 
to the thin disk, which uses
several fields to analyse the data using a synthetic model. Recently, 
Buser et al. (1999) find the Galactic thick disk to have local density 
of 5.9$\pm$3 \% of the local thin disk density and exponential scale height 
of h$_{z}$ = 910$\pm$300 pc, using the new Basel RGU high-latitude survey. 

\begin{table*}
\caption[]{The best fitted structural parameters of the thin and thick
disk stars with M$_{V}$$\ge$3.5 derived from the GAR field. The mean
structural parameters derived from galactic anticentre and centre fields 
are also presented in the table}
\begin{flushleft}
\normalsize
\begin{tabular}{c c c c}\\
\hline
Field & Thin Disk & Thick Disk & Thin Disk : Thick Disk\\
\hline
3.5$\le$M$_{V}$$\le$6 & h$_{z}$ (pc) & h$_{z}$ (pc) & density ratio\\
\hline
GAR    & 240$\pm$20  & 790$\pm$10  & 100 : 6.1$\pm$3.0\\
GAC1,2 & 301$\pm$35  & 828$\pm$21  & 100 : 8.0$\pm$0.7\\
GC     & 222$\pm$4   & 700$\pm$24  & 100 : 9.8$\pm$0.1\\
\hline
\end{tabular}
\end{flushleft}
\end{table*}

\section{Kinematics of thick disk population}

The cardinal components of the stellar space velocity (in km/s), U, V
and W were derived from proper motions, $\mu_{l}$ and 
$\mu_{b}$ (in arcsec year$^{-1}$) and distance d (in pc). 
In the direction of GAR field, we are measuring the
velocities (U, V-W) defined as (see more detail in Paper III) :

$U \simeq 4.74 ~d ~\mu_{l}cos b$ ~~~~~~~and~~~~~~~~  ${V-W \over \sqrt{2}} \simeq 4.74 ~d ~\mu_{b}$

and we define : 

$\sigma^{2}_{V,W} = {\sigma^{2}_{V} + \sigma^{2}_{W} \over 2}$

To perform the kinematical separation in our sub-sample of stars
(0.3$<$B-V$<$0.9), we have used a maximum likelihood method 
(SEM algorithm : Celeux \& Diebolt 1986) in order to deconvolve the
multivariate gaussian distributions and estimate the corresponding 
parameters. The aim of the SEM algorithm is to resolve the finite 
mixture density estimation problem under the maximum likelihood 
approach using a probabilistic teacher step. This method has already 
been used by Soubiran (1993ab) and in Paper I, II \& III
to characterize the (U,V,W) parameters of the
stellar populations.

The samples of stars in GAR field have been devided in several bins of
distance, and in each bin of distance, a fit has been performed with a 
SEM algorithm to separate the 2-D gaussian distributions to identify 
the three components (thin disk, thick disk and halo) of the Galaxy.
In Fig. 5, the three gaussian populations representing the thin disk,
the thick disk and the halo are overplotted on the (V-W)/$\sqrt{2}$ 
velocity histogram for the distance interval 1000$\le$d$\le$1500 pc.
The mean kinematic parameters of thick disk derived from GAR field,
up to the distance of 3.5 kpc above the galactic plane is shown in Table 9.
In the same table, we have also presented the kinematics of the thick
disk population obtained from the galactic anticentre and centre fields.

\begin{figure}
\psfig{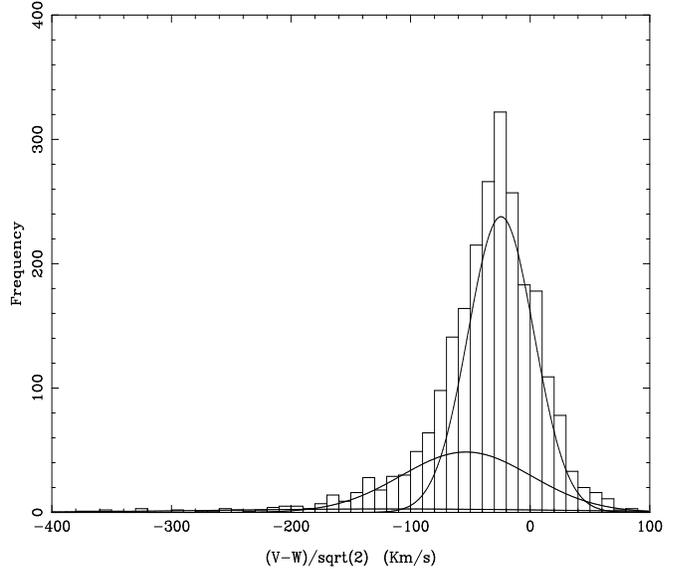}
\caption[]{Histogram of the $V-W \over \sqrt{2}$ for stars in distance
interval 1000$\le$d$\le$1500 pc towards GAR direction. The 3 gaussian 
components solution of SEM corresponding to thin disk, thick disk, and
halo are overplotted}
\end{figure}

\begin{table*}
\caption[]{The mean kinematic parameters of thick disk (in km/s) derived
from GAR field. The kinematic parameters derived from galactic
anticentre and centre fields are also presented in the table}
\begin{flushleft}
\normalsize
\begin{tabular}{c c c}\\
\hline
Antirotation (GAR) & Anticentre (GAC1,2) & Centre (GC)\\
\hline
$\sigma_{U}$ = 67$\pm$12 & $\sigma_{U,W}$ = 64$\pm$4 & $\sigma_{U,W}$ = 66$\pm$3\\
$\sigma_{V,W}$ = 51$\pm$12 & $\sigma_{V}$ = 60$\pm$3 & $\sigma_{V}$ = 56$\pm$2\\
{$<(V-W)> \over \sqrt{2}$} = -49$\pm$10 & $<V>$ = -57$\pm$4 & $<V>$ = -49$\pm$3\\
\hline
\end{tabular}
\end{flushleft}
\end{table*}

The thick disk parameters derived from GAR field are in good agreement 
with others : Soubiran 1993ab; Beers \& Larsen 1994; Bartasiute 1994 and 
Paper III (derived from 2 intermediate latitude fields). By 
combining the four fields, GAC1,2, GC, GAR and NGP (Soubiran 1993ab), we have
derived the mean kinematical parameters of thick disk (as shown in table 5
in Paper III), which are : $\sigma_{U}$ = 67$\pm$4 km/s, 
$\sigma_{V}$ = 51$\pm$3 km/s, $\sigma_{W}$ = 40 km/s and V$_{Lag}$ = 
-53$\pm$10 km/s (with respect to the Sun). We do not find any evidence for 
a vertical gradient in the thick disk kinematics in GAR field, in agreement 
with our previous results in the anticentre and centre fields, and with the 
field near the pole (Soubiran 1993ab). The suggested gradient from Majewski 
(1992) is not confirmed. As shown in table 6 in Paper III that since 5-6 
years, the accuracy of the kinematical measurements of the thick disk 
population has greatly improved and the results are well in agreement with
each others.

\begin{figure*}
\psfig{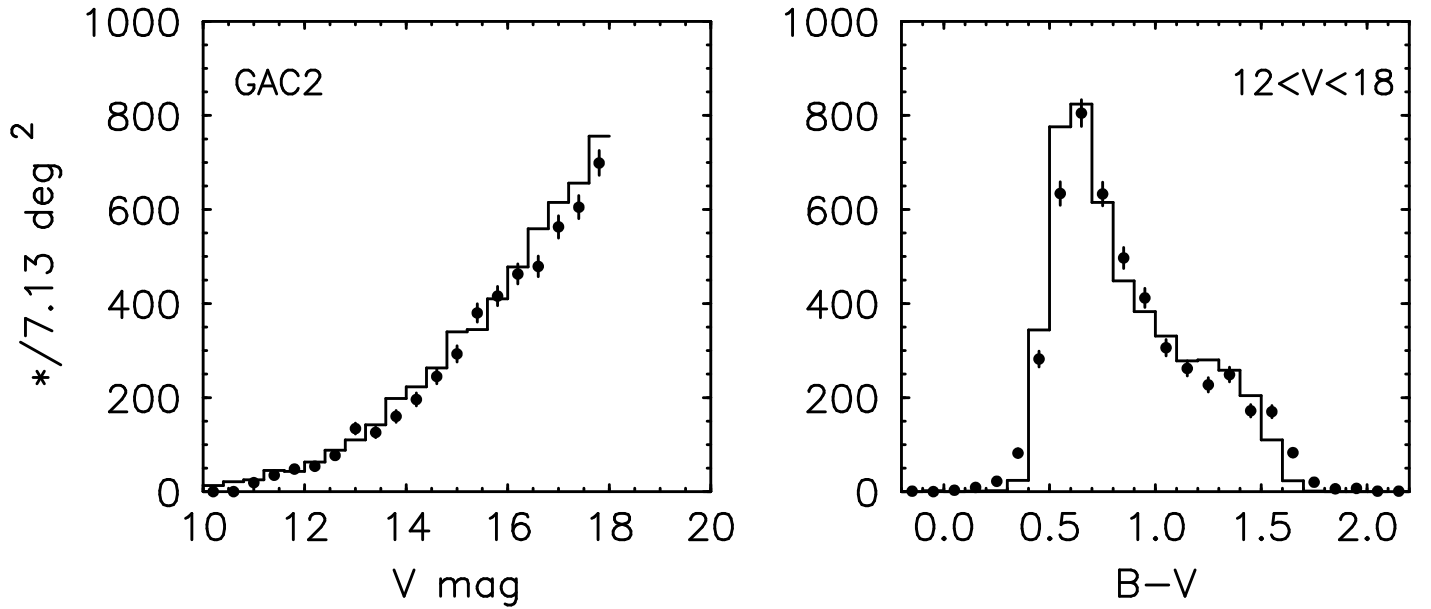}
\caption[]{Comparison of V star counts and B-V color between the observed 
data sets and the Besan\c con model predictions towards GAC2 direction.
The histogram represents the model predictions. The filled circles
are the observed data}
\end{figure*}

\begin{figure*}
\psfig{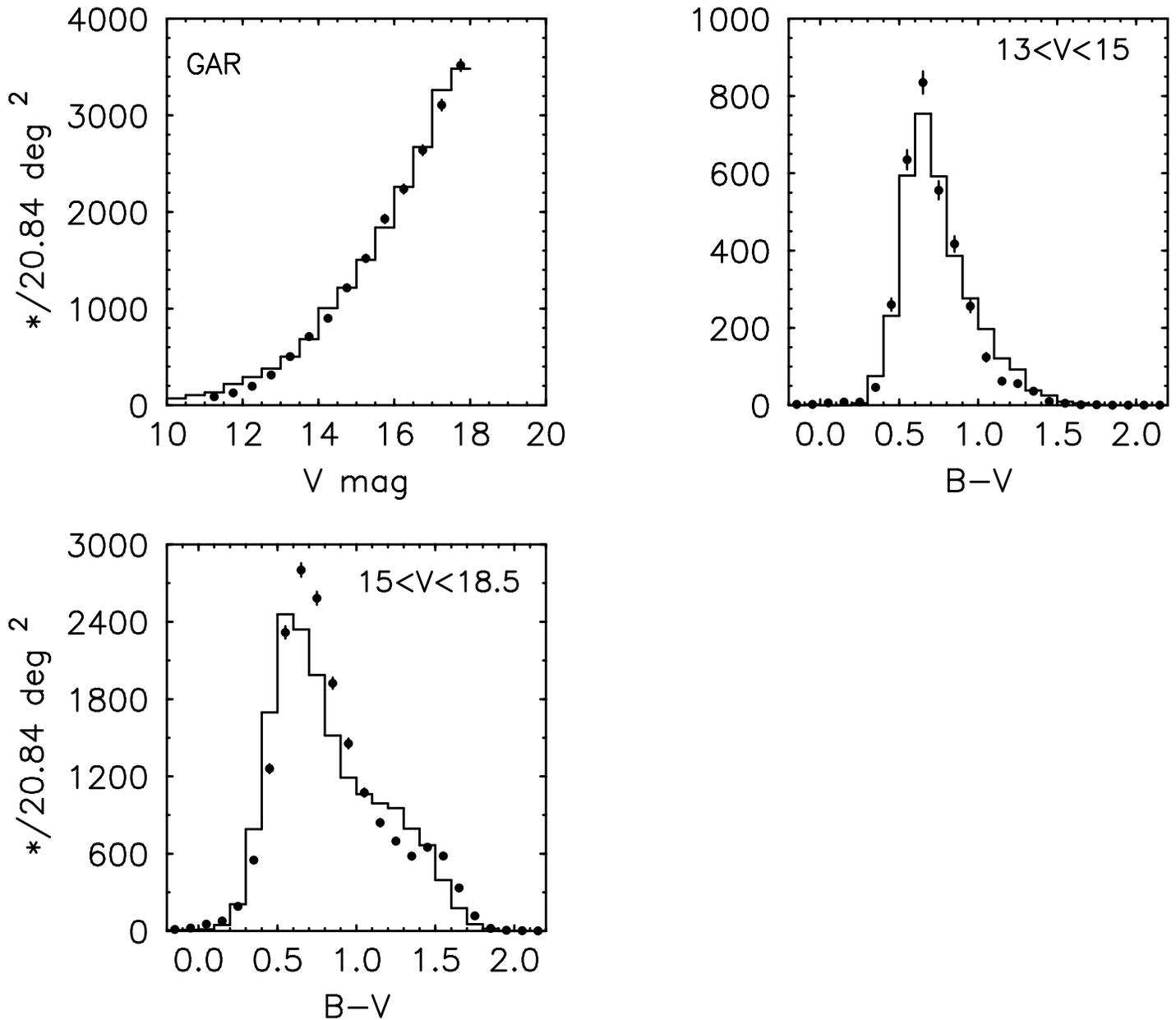}
\caption[]{Comparison between the observed V star counts \& B-V color 
distributions and the model predictions towards GAR direction. The symbols 
are as in Fig. 6}
\end{figure*}

\begin{figure*}
\psfig{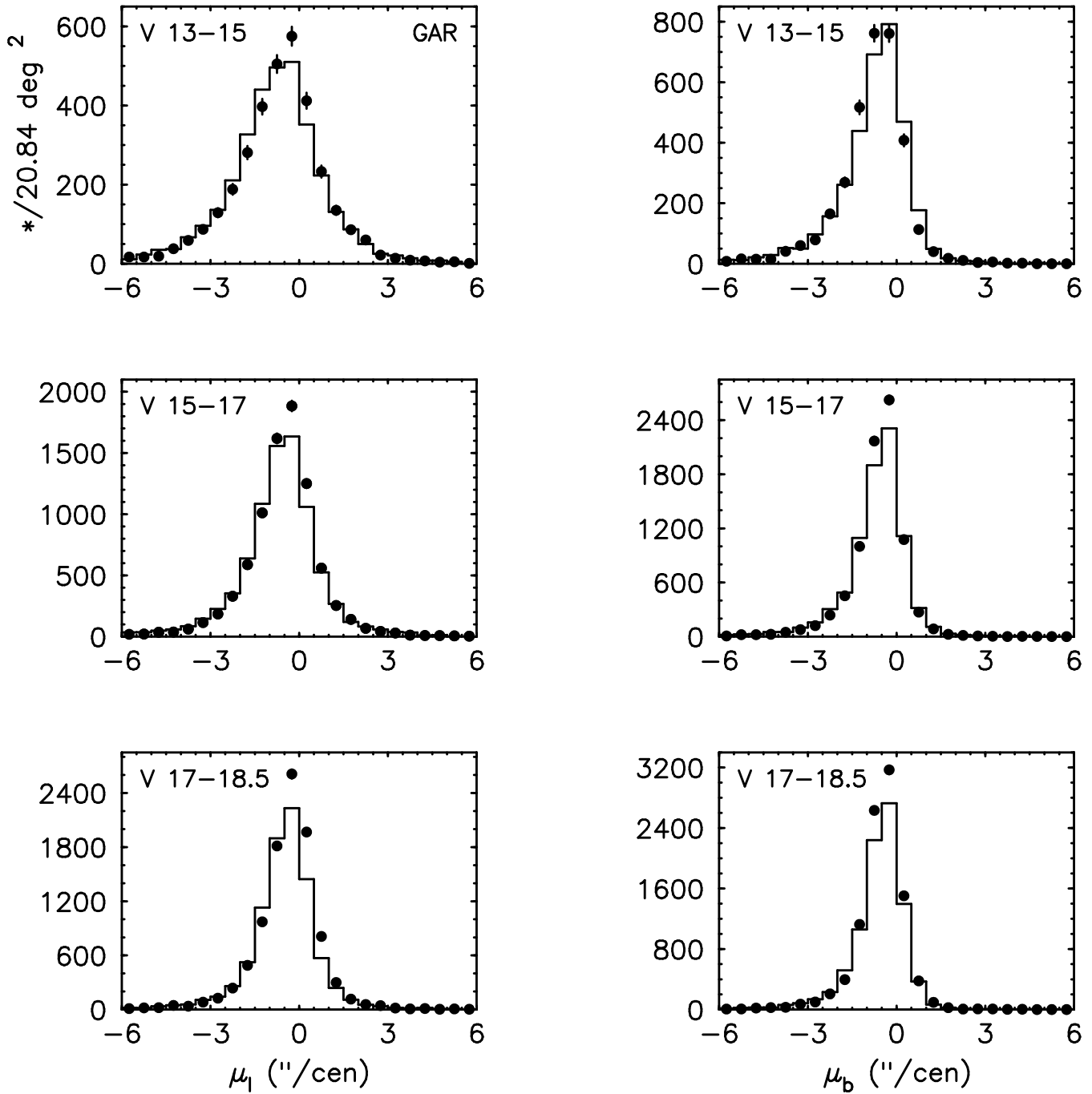}
\caption[]{Comparison of $\mu_{l}$ \& $\mu_{b}$ distributions
with the model predictions towards GAR direction. The symbols are
as in Fig. 6}
\end{figure*}

\section{Comparison of our data sets with Besan\c con model of stellar 
population synthesis}

\subsection{Besan\c con model}

The data have been compared to the Besan\c con model (Robin \& Cr\'ez\'e 1986, 
1996; Bienaym\'e et al. 1987; Haywood 1994, 1997), by simulating the
distributions of magnitudes, colors and proper motions including the
photometric and astrometric errors as determined. The Besan\c con model
of stellar population synthesis has been developed for about 13 years as
an attempt to put together all constraints (theoretical and observational)
about galactic evolution in order to obtain a consistent scenario of galaxy
evolution. We used the new version of the model (Robin et al. 1999).
To minimize the Poisson 
noise in model simulations, the catalogues are simulated 5 or 10 times the 
total area of the observed data set. While the simulations in the two fields 
are qualitatively in agreement with the data, some discrepancies exist. In 
particular, the model overpredicts the number of stars in the GAC2 field  
beyond V$>$17 (Fig. 6). In GAR field, the number of stars are also slightly 
overpredicted at the bright magnitudes (V$<$12) (Fig. 7). All together, the 
models seem to predict slightly more red disk stars than visible in the data
(B-V$\sim$1.0 - 1.2). The comparison between the model predictions and the 
observed magnitude, color and proper motion distributions in two galactic 
fields is shown in Figs. 6, 7 \& 8. 

\subsection{GAC2 field ($l=167.5^\circ$, $b=47.4^\circ$)}

The distribution of the observed star counts in V band is shown in Fig. 6.
In the same diagram, the model predicted counts are overplotted. The
error bars are $\pm\sqrt{N}$, where N is the number of stars in each bin.
The two distributions are in good agreement in the magnitude range 
11$<$V$<$17, however, 
we find that the model predicts slightly more stars beyond V = 17
towards this direction. The possible reasons of this discrepancy between
the model and data are discussed in \S 8.

The comparison between the observed data set and the Besan\c con model
predictions in B-V color indice is also shown in Fig. 6. There is a good 
agreement between the two distributions in the magnitude range 
12$\le$V$\le$18. At B-V $\sim$ 1.2 (in Fig. 6), the more red disk stars are 
seen in the model compared to the data.

\subsection{GAR field ($l=277^\circ$, $b=47^\circ$)}

Fig. 7 shows the observed V distribution towards GAR field. We compare 
these observations with the predictions of the Besan\c con model overplotted
on the same figure. The total observed counts are in good agreement with the
model predictions in the magnitude range 12$<$V$<$18, however, the contribution
of blue stars at V$<$12 is slightly larger in the model.

The comparison between the observed distributions and the model predictions 
in B-V color indice is also shown in Fig. 7. The two distributions in B-V 
color are in good agreement in the magnitude range 13$<$V$<$15. In the 
magnitude range 15$<$V$<$18.5, the model overpredicts the red disk stars 
(B-V $>$1.0).

Model predicted proper motions distributions are compared to the observed ones
in Fig. 8, assuming the same observational errors. The proper
motion distributions (in $\mu_{l}$ \& $\mu_{b}$) agree well for stars
in all the magnitude ranges. The mean velocity ellipsoid of the
thick disk population used in the model simulation is the one deduced from 
our previous investigations in intermediate latitude fields and from the 
present paper. Due to the stellar drift, the asymmetric shape of the
distribution of latitude proper motion is clearly visible in data and model.
Proper motion distribution is sensitive to the asymmetric drift and the
velocity dispersions of the stellar populations. The comparison between 
the data and model distributions in Fig. 8 is quite well, particularly at
15$<$V$<$17, where the majority of thick disk stars exist. It also proves
that our value of the thick disk velocity ellipsoid is well determined.  

\subsection{Discussion}

We have matched our data sets against the Besan\c con model of population 
synthesis. The main aim of these comparisons is to check the coherence of the 
model. Although, the total number counts brighter than V=18.5 are in relatively 
good agreement with observations, the contribution of blue stars of disk 
(V$<$12) is slightly larger in the model in GAR direction. The reason of this 
discrepancy is difficult to trace back with the present data. It could be 
attributed to two different parameters in the model, or a mixed of these two : 
the HR diagram (e.g the relative distribution of disk stars in the HR diagram) 
or the normalisation adopted for the disk stars, the 2nd parameters being the 
most critical. Since there is almost no data available to constraint the model 
between magnitudes 9-12, the model is almost entirely defined by its fit to 
the solar neighborhood luminosity function (See Haywood 1994, 
Haywood et al. 1997). The luminosity function (LF) adopted to normalize the 
model was that published by Wielen et al. (1983), and it is known to 
overestimate the number of stars in the 
25 pc sphere by almost 30\% (Turon, 1996). This problem could also be 
attributed to the distribution of the disk stars in the HR diagram, in 
particular, since the B-V color at bright magnitudes (V=9-12) are 
uncorrectly shifted to the red. However, simulations made with different SFR 
for disk stars, changing the relative distribution in the HR diagram, did not 
improve significantly the B-V model-data comparisons. 
One expects that these two aspects will be much better known now after the 
analysis of the Hipparcos and Tycho catalogues, which provides strong 
contraints on the distribution of the stars at V$<$10.5. Besan\c con model
is in further improvement phase by using the data from Hipparcos and
Tycho catalogues.

At fainter apparent magnitudes, where the disk becomes less important 
(V$>$12), the predictions are in good agreement in B-V (Figs. 6 \& 7). 
At V$>$12, the absolute magnitude of the dominant stars is not the same
that at V$<$12.
However, at V=16-18, there are again slightly more stars in the model than 
in the data, in the two fields at B-V $\sim$ 1.2. This can not be attributed 
to incompletness in the data, since the catalogues are completed 
up to V=18.5 and B=20. According to galactic models, red stars (B-V$>$1) at 
these magnitudes (V$>$17) are disk stars, whereas at B-V$\ge$1.2 and 
16$<$V$<$17, there is a number of thick disk stars which could eventually make
the overprediction of the model. This discrepancy could be due the
luminosity function of the thick disk which may be inappropriate, if this
is not the density law. If the present discrepancy is significant, 
one could possibly be attributed to unresolved binaries also, which are not 
accounted for in the model.

\section{Conclusions}

We have presented a new survey of absolute stellar proper motions and 
multicolor photometry to V = 18.5 in two directions at intermediate
latitude. The most probable value of scale height for the thick disk
stars in GAR direction is determined to be h$_{z}$$\simeq$790$\pm$10 pc 
with a local density of 6.1$\pm$3 \% relative to the thin disk. The 
velocity ellipsoid of the thick disk component has been determined. These
values are in perfect agreement with the recent determination given
by Robin et al. (1996) and Ojha et al. (1996) based on the analysis
of a large set of available photometric and astrometric catalogues. The
Besan\c con model predictions are also compared with the present survey in
the overlapping magnitude ranges.

\acknowledgements

This research was partially supported by the Indo-French Center for the
Promotion of Advanced Research / Centre Franco-Indien Pour la
Promotion de la Recherche Avanc\'ee, New Delhi (India). We thank all the 
MAMA, ESO, Tautenburg and Leiden Observatory staffs who made this 
investigation possible. We especially thank referee Dr. Gerry Gilmore
for his comments. We also thank Misha Haywood for going through an
earlier version of this manuscript and giving his useful comments.
This research has made use of the DEC-ALPHA system of the Optical CCD 
astronomy programme of TIFR.

\end{document}